\newcommand{\Jpsi}{\rm J/$\psi$ }
\newcommand{\state}{$\Upsilon$(1S) }

\newcommand{\raa}{$R_{\rm AA}$ }

\documentclass[twocolumn]{article}

\usepackage{rotating}
\usepackage{textcomp}
\usepackage{multirow}
\usepackage{lineno}
\begin{document}

\title{Quarkonium measurements in Pb\--Pb and p\--Pb collisions with \hbox{ALICE} at the LHC}

\author{Lo\"ic Manceau (lmanceau@to.infn.it) for the ALICE Collaboration\\ INFN sezione di Torino\\Proceddings for LHCP 2013 international conference}

%
%

\maketitle

\abstract{
The ALICE experiment has measured quarkonium production in \hbox{Pb\--Pb} collisions at $\sqrt{s_{\rm NN}}=2.76$ TeV and in p\--Pb collisions at $\sqrt{s_{\rm NN}}=5.02$ TeV. The measurements are performed in the dielectron decay channel at mid-rapidity ($|y_{\rm LAB}| < 0.9$) and in the dimuon decay channel at forward rapidity ($2.5 < y_{\rm LAB} < 4$). We focus in particular on the \Jpsi nuclear modification factor measured both in Pb\--Pb and p\--Pb collisions and on the \state nuclear modification factor measured in Pb\--Pb collisions. 
}
\section{Introduction}
\label{intro}
The primary aim of ultra-relativistic heavy-ion collisions is to produce nuclear matter at high temperature. Under this conditions Quantum Chromodynamics predicts the existence of a deconfined state of partonic matter or QGP (Quark-Gluon Plasma). Heavy quarks are produced in hard scatterings occurring in the early stages of the collision before the deconfinement phase and interact with the medium. Quarkonium measurements thus provide essential information on the medium. In addition, quarkonium production in AA collisions is expected to be sensitive to the nuclear modification of parton distribution functions (shadowing and anti-shadowing classified as nuclear initial state effects) and to the quarkonia energy loss or break up in cold nuclear matter (nuclear final state effects). Measuring quarkonium production in pA where no QGP is expected, should help to quantify nuclear initial/final state effects.

According to the color-screening model~\cite{Matsui:1986dk}, the measurement of the dissociation probability of the different quarkonium states is expected to provide an estimate of the initial temperature of the QGP. In particular, the $\Upsilon$(1S) should dissociate at much higher temperature than all the other bottomonium (or charmonium) states. Together  with the other $\Upsilon$ resonances, bottomonia can therefore be considered as a very effective thermometer of the system~\cite{Digal:2001ue}. Extensive experimental results on J/$\psi$ production in AA collisions at SPS~\cite{Kluberg:2005yh} and RHIC~\cite{Levy} show a significant suppression with respect to pp and even pA collisions. However, the interpretation of RHIC results is not so straightforward and have lead to the development of various models. A class of models predict a significant contribution to quarkonium production due to the in medium recombination of un-correlated heavy quark pairs~\cite{Thermo,Zhao:2011cv,Liu:2009nb}. The increasing production rate of the latter with increasing energy should make the AA collisions at the LHC suitable for detecting possible enhancement effects on the \Jpsi production. In what concern bottomonium production by recombination, it should be much less important as the initial production rate for $b\bar{b}$ pairs is smaller than that of $c\bar{c}$. 

ALICE (A Large Ion Collider Experiment)~\cite{KAaJINST} is the LHC experiment dedicated to the study of heavy-ion collisions. At forward rapidity ($2.5<y_{\rm LAB}<4$), ALICE is equipped with a spectrometer which allows to measure quarkonia via their dimuon decay channel. 
At mid-rapidity ($|y_{\rm LAB}|<0.9$), the ALICE central barrel makes possible the measurement of quarkonia via their dielectron decay channel. 
In Pb\--Pb collisions, events are selected according to their degree of centrality by means of the VZERO hodoscope made of two scintillator arrays covering the pseudorapidity ranges $2.8<\eta<5.1$ and $-3.7<\eta<-1.7$. The VZERO also contributes to the Minimum-Bias (MB) interaction trigger decision.

\section{Data Analysis}
\label{sec-1}


Table~\ref{tab-1} summarises the integrated luminosity of the analysed data sample, the energy of the collision and the rapidity range for each of the measurements presented in these proceedings.
While mid-rapidity data were collected with a MB trigger, forward rapidity data are enriched with unlike-sign dimuons by means of the muon spectrometer trigger system. 
\begin{center}
\begin{table*}[ht]
\centering
\caption{Summary of the analysis presented in these proceedings and integrated luminosity of the used data samples.}
\label{tab-1}       
\begin{tabular}{llllll}
\hline
system & Pb\--Pb & Pb\--Pb & p\--Pb & Pb\--p & Pb\--Pb \\\hline
Energy ($\sqrt{s_{\rm NN}}$)& $2.76$ TeV & $2.76$ TeV & $5.02$ TeV & $5.02$ TeV & $2.76$ TeV \\
$y_{\rm LAB}$ & $2.5<y<4$ & $|y|<0.9$ & $2.5<y<4$ & $2.5<y<4$& $2.5<y<4$\\
measured particle &\Jpsi &\Jpsi &\Jpsi & \Jpsi & \state\\
Integrated luminosity & $69.5\ \mu$b$^{-1}$ & $\sim 15\ \mu$b$^{-1}$ & $4.90$ nb$^{-1}$ & $4.49$ nb$^{-1}$  & $69.2\ \mu$b$^{-1}$ \\\hline
\end{tabular}
\end{table*}
\end{center}
The relative production of quarkonia in collisions involving a Pb nucleus with respect to that in pp collisions, can be quantified with the nuclear modification factor ($R_{\rm AA}$). The $R_{\rm AA}$ can be expressed in term of the quarkonium invariant yield ($Y$), the nuclear overlap function ($\langle T_{\rm AA}\rangle$) and the quarkonium pp cross section at the same energy in the nucleon-nucleon center of mass ($\sigma^{\rm pp}$) as follows:

\begin{equation}
R_{\rm AA}=\frac{Y}{\langle T_{\rm AA}\rangle\sigma^{\rm pp}}.
\label{RAAFormula}
\end{equation}

The invariant yield ($Y$) corresponds to the probability for a quarkonium state to be produced in the detector acceptance during a MB event. The estimate of such a quantity needs correcting the measured number of dileptons from quarkonium decays for the acceptance and the efficiency of the detector and for quarkonia-to-dilepton branching ratio. The nuclear overlap function is equal to the average number of binary collisions divided by the nucleon-nucleon cross-section and can be interpreted as the nucleon-nucleon equivalent integrated luminosity per hadronic collision at a given event centrality~\cite{TAAREF}. In the case of the \Jpsi production measurement in Pb\--Pb collisions the pp reference cross section was directly measured with ALICE~\cite{Abelev:2012kr}. For other studies presented in these proceedings, the pp reference cross section was obtained by a phenomenological extrapolation of available data on quarkonium hadroproduction at various energies~\cite{ZaidaInter,UpsiPbPbPN}.

\section{\Jpsi production in Pb\--Pb}
\label{sec-2}
\begin{figure}
\centering
\includegraphics[width=8.25cm]{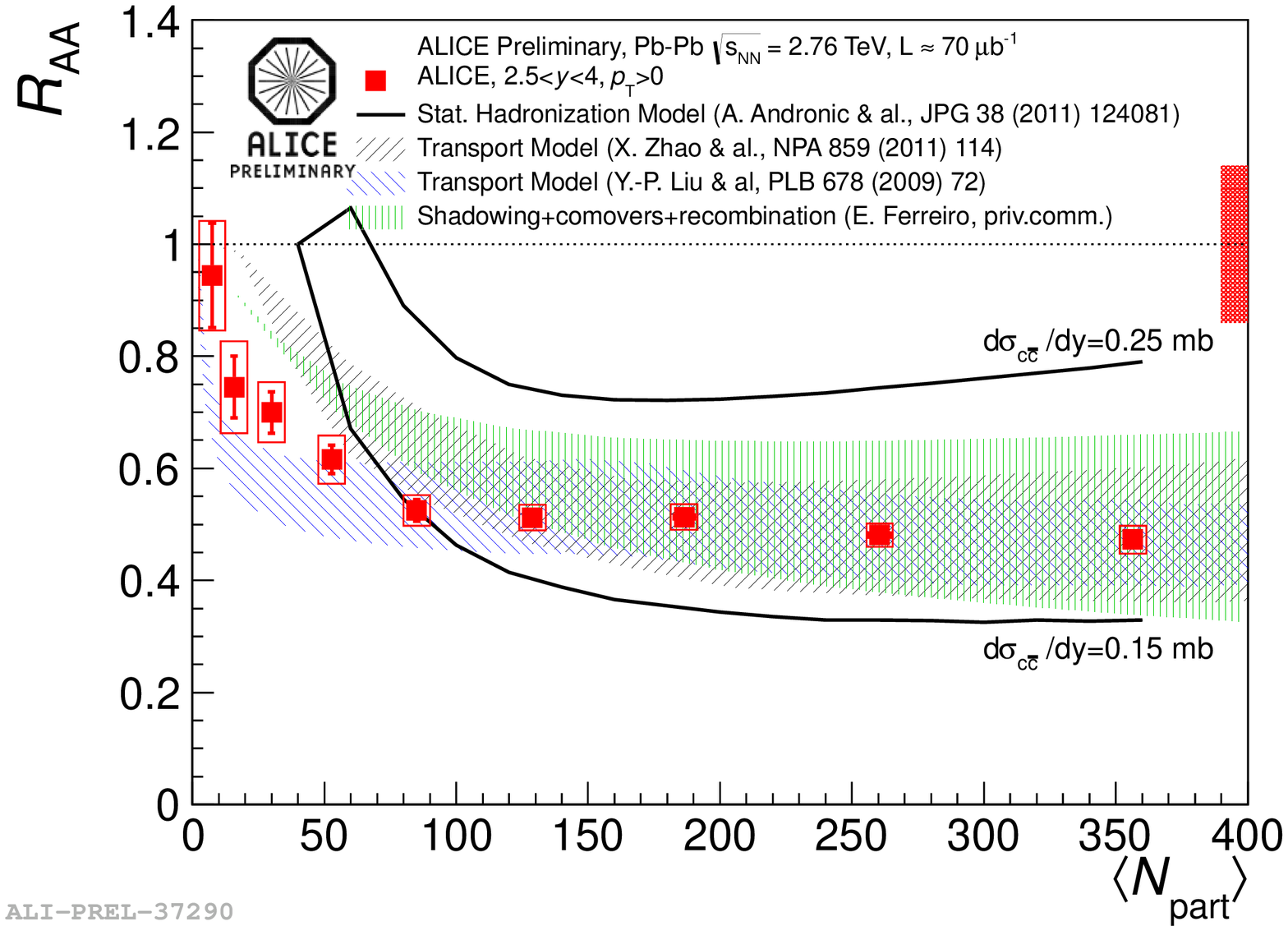}
\includegraphics[width=8.25cm]{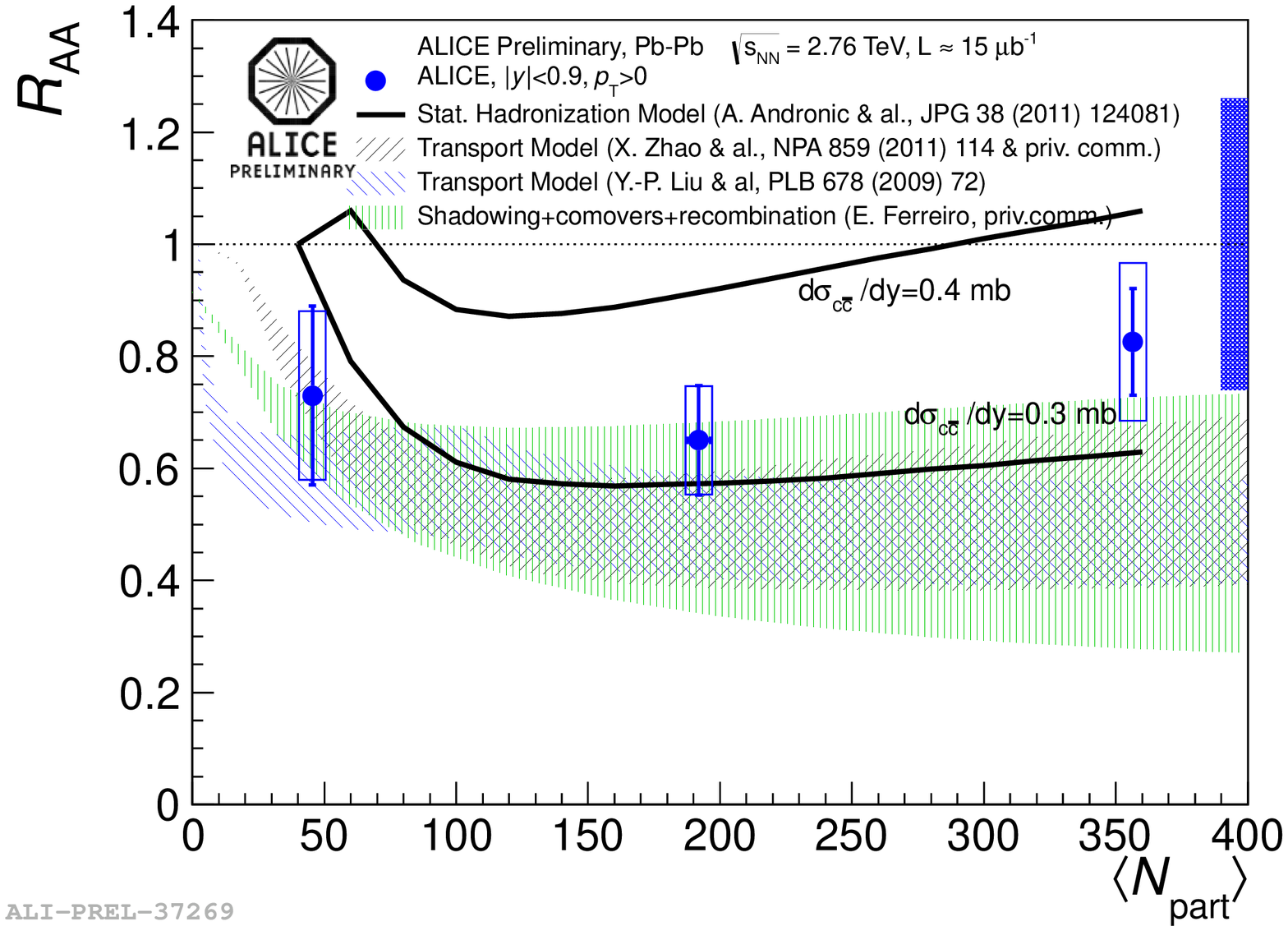}
\caption{\Jpsi nuclear modification factor as a function of the average number of participant nucleons in Pb\--Pb collisions at $\sqrt{s_{\rm NN}}=2.76$ TeV measured at forward rapidity (left) and at mid-rapidity (right). Data is compared to different models including a large contribution from (re-)generated \Jpsi.}
\label{fig-1}       
\end{figure}

The \Jpsi $R_{\rm AA}$ measured by ALICE in~Pb\--Pb collisons at $\sqrt{s_{\mathrm{NN}}} = 2.76$ TeV in the kinematic range $2.5 < y_{\rm LAB} < 4$ ($|y_{\rm LAB}|<0.9$) and $\ensuremath{p_{\rm T}}>0$ GeV/c is shown as a function of the number of participant nucleons in the left (right) panel of Fig.~\ref{fig-1}. There is a clear suppression which does not exhibit a significant centrality dependence for $\langle N_{\rm part}\rangle>100$. At RHIC energies, the centrality dependence is more pronounced~\cite{Adare:2011yf,Adare:2006ns}.

ALICE measurements are compared with theoretical models that include a \Jpsi (re-)generation component from deconfined charm quarks in the medium. The Statistical Hadronization Model~\cite{Thermo} assumes deconfinement and a thermal equilibration of the bulk of the ${\rm c\bar{c}}$ pairs. Then charmonium production occurs at phase boundary by statistical hadronization of charm quarks. The prediction is given for two estimated values of ${\rm d}\sigma_{\rm c\bar{c}}/{\rm d}y_{\rm LAB}$, since a measurement of this quantity still does not exist for Pb\--Pb collisions. The two transport model results~\cite{Zhao:2011cv,Liu:2009nb} presented in the same figures differ mostly in the rate equation controlling the \Jpsi dissociation and regeneration. Both are shown as a band which connects the results obtained with (lower limit) and without (higher limit) shadowing. The width of the band can be interpreted as the uncertainty of the prediction. In both transport models, the amount of regenerated \Jpsi in the most central collisions contributes to about $50\%$ of the measured yield, the rest being from initial production.

\section{\Jpsi production in p\--Pb}
\label{sec-3}

\begin{figure}
\centering
\includegraphics[width=8.25cm]{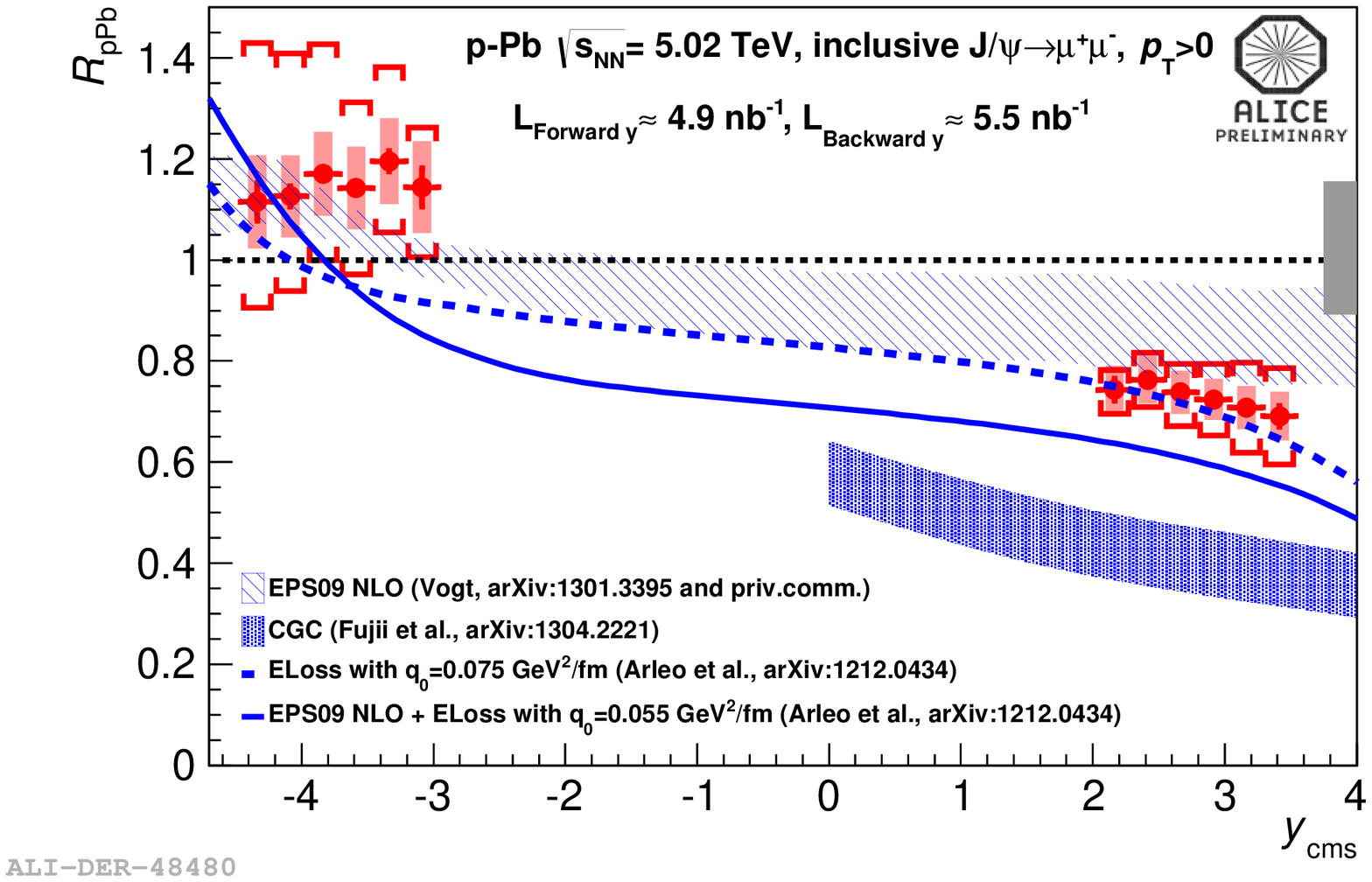}
\includegraphics[width=8.25cm]{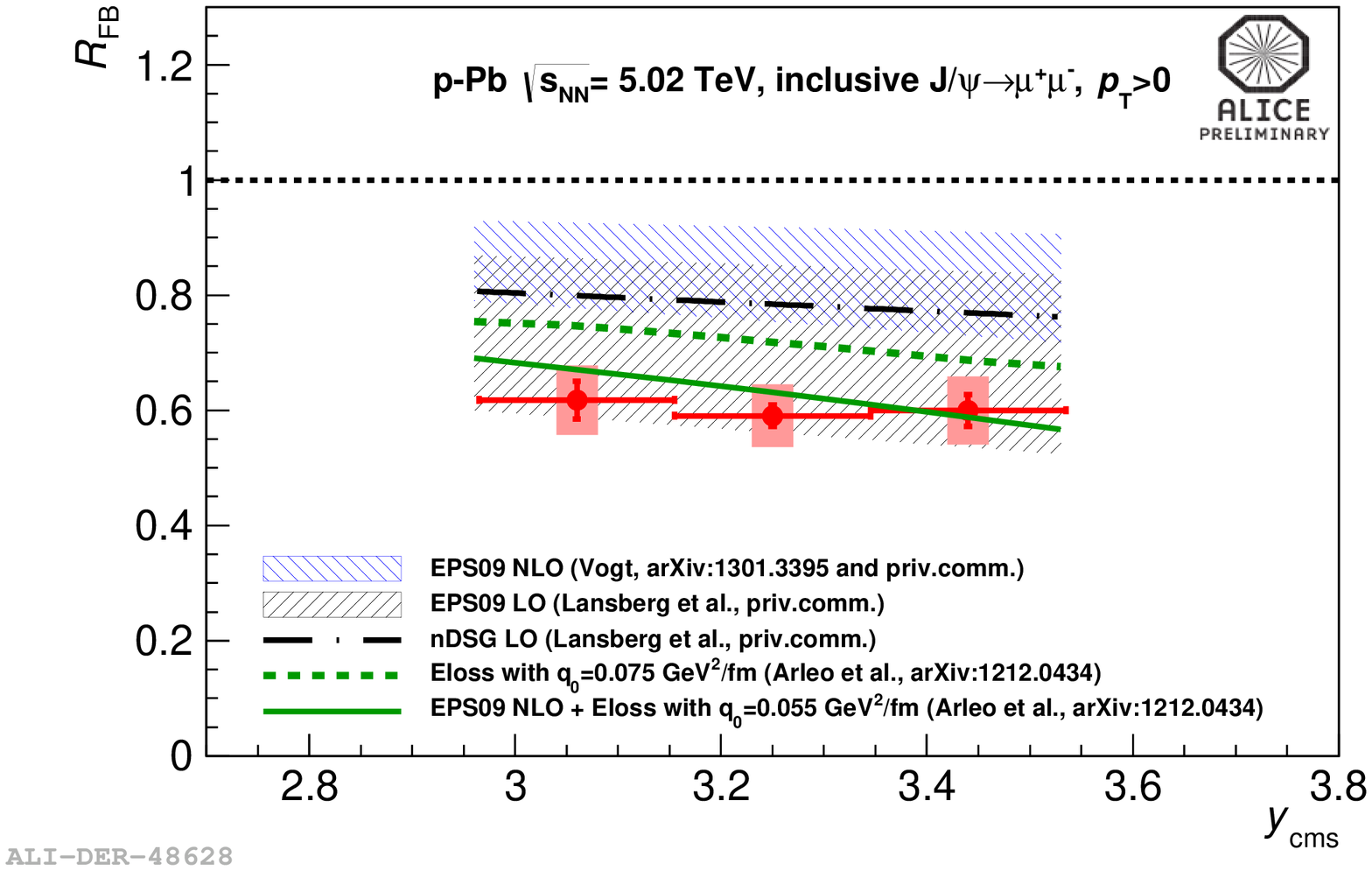}
\caption{Left: \Jpsi nuclear modification factor as a function of the rapidity in the center of mass of the p\--Pb/Pb\--p system. Right: Ratio of the nuclear modification factor at forward rapidity by that at backward rapidity as a function of the rapidity in the center of mass of the p\--Pb/Pb\--p system. Data are compared to models describing different nuclear initial state effects. It is worth underlining that $\sqrt{s_{\rm NN}}=5.02$ TeV.}
\label{fig-2}       
\end{figure}

The \Jpsi nuclear modification factor was measured down to $p_{\rm T}=0$ as a function of the center of mass rapidity ($y_{\rm CM}$) for p\--Pb collisions at $\sqrt{s_{\rm NN}}=5.02$ TeV (Fig.~\ref{fig-2} (left)). Two beam configurations were studied, corresponding to Pb nuclei moving away from (p\--Pb) or towards (Pb\--p) the muon spectrometer. In this way one can access $R_{\rm pA}$ at forward and backward rapidity, respectively. As the energy of the two beams is not the same, the detector rapidity coverage in the laboratory frame ($2.5<y_{LAB}<4$) corresponds to different coverage in the system center of mass, $2.03<y_{CM}<3.54$ (p\--Pb), and $-4.46<y_{CM}<-2.96$ (Pb\--p). A suppression increasing from backward to forward rapidity is observed.

The ratio ($R_{\rm FB}$) between the nuclear modification factors measured in the common rapidity range acessed at forward and backward rapidity ($2.96<|y_{\rm CM}|<3.54$) is shown (Fig.~\ref{fig-2} (right)) as a function of $y_{\rm CM}$. The $R_{\rm FB}$ exhibits a small $y_{\rm CM}$ dependence and is about $0.6$.

Data were compared to EPS09 calculations at Next-to-Leading-Order~\cite{Albacete:2013ei}, to a model of coherent parton energy loss~\cite{Elossref} implemented with or without shadowing (EPS09 at Next-to-Leading-Order) and to a low-$x$ saturation model (Color Glass Condensate)~\cite{CGCref}. Among the considered nuclear initial state effect models, EPS09 at Next-to-Leading-Order is found to reproduce reasonably the data with or without the consideration of coherent parton energy loss while the low-$x$ saturation model is not favoured.   

\section{\state production in Pb\--Pb}
\label{sec-4}

\begin{figure}
\centering
\includegraphics[width=8.25cm]{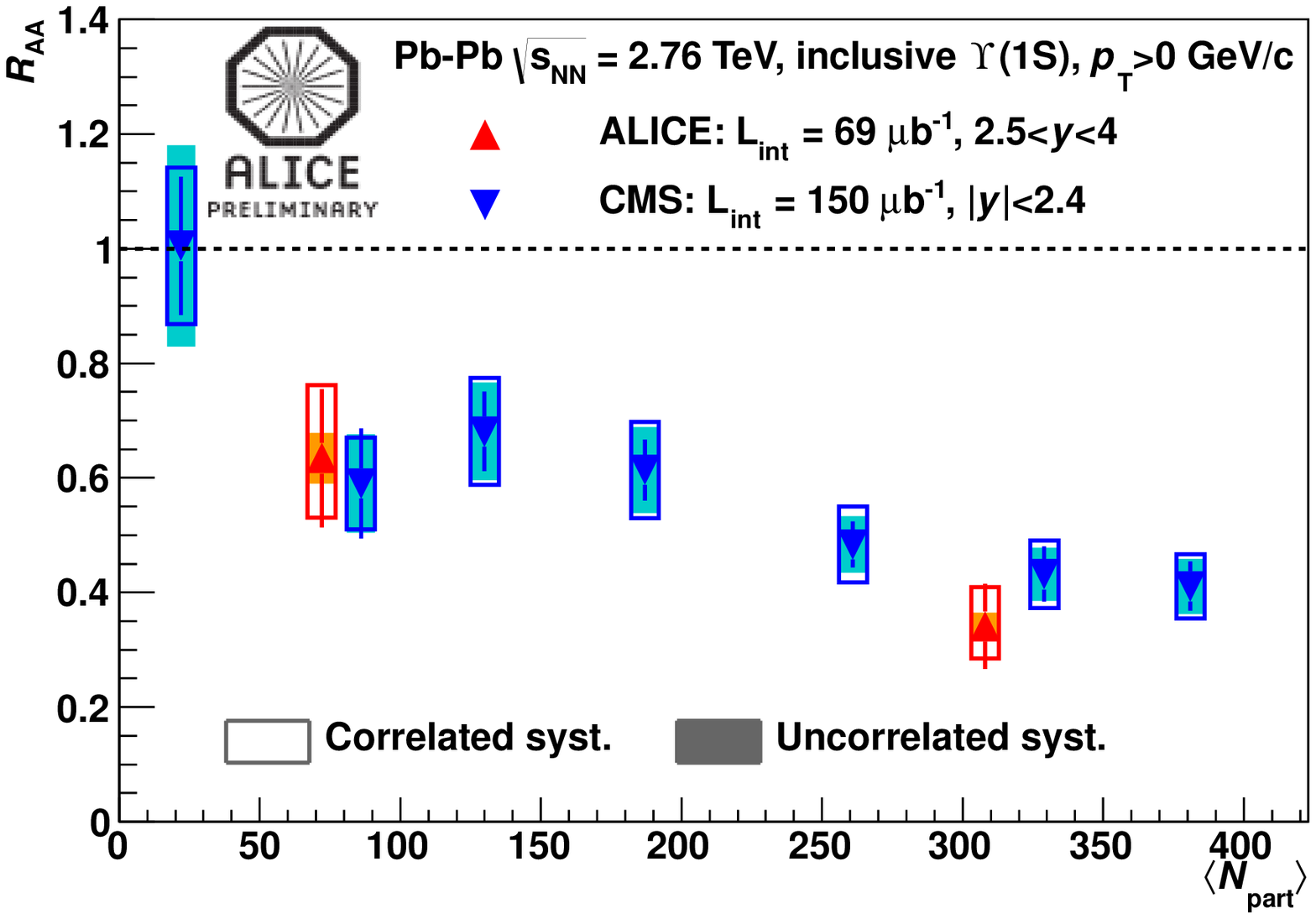}
\includegraphics[width=8.25cm]{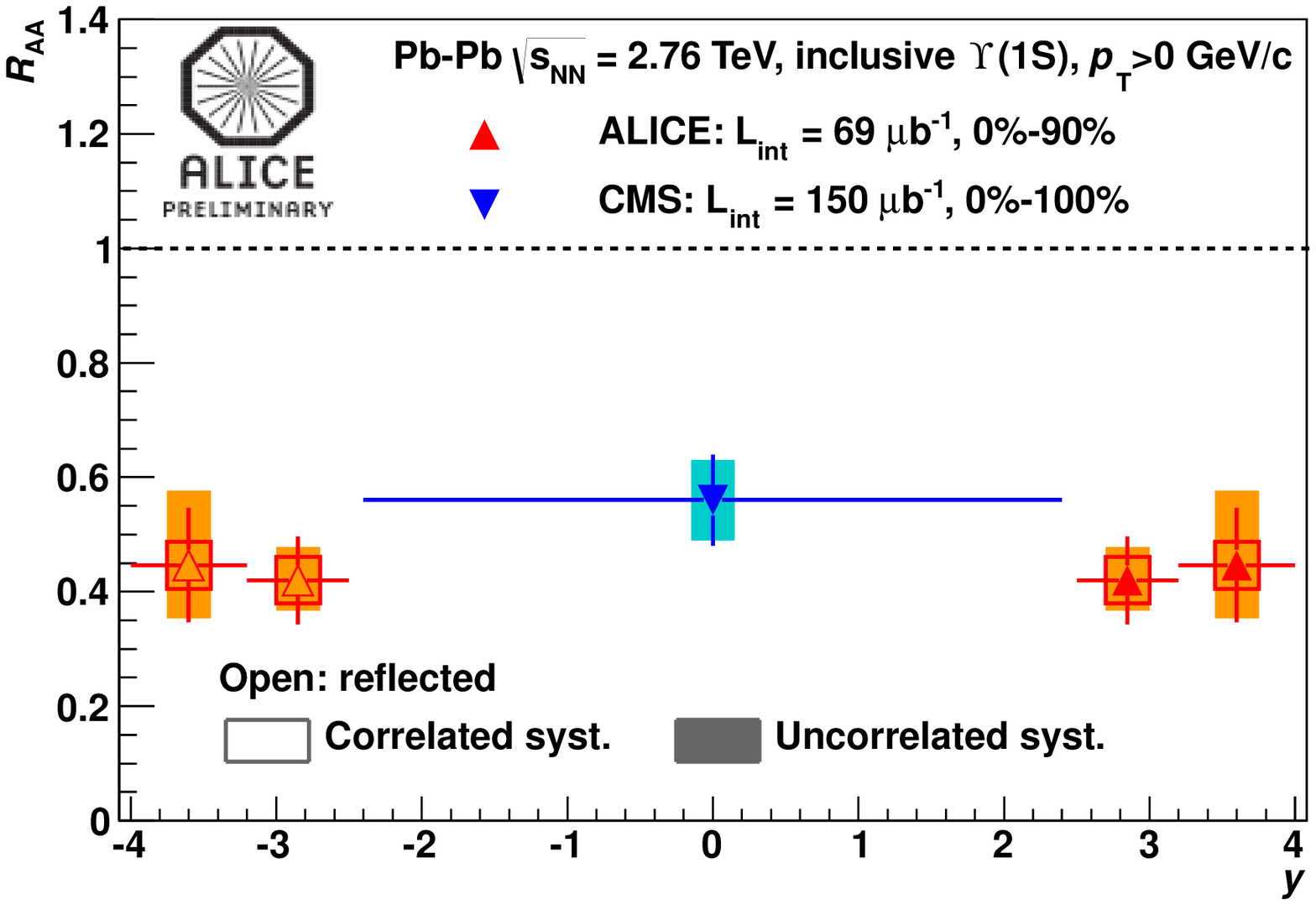}
\caption{Left (Right): ALICE~\cite{UpsiPbPbPN} and CMS~\cite{Chatrchyan:2012lxa} nuclear modification factor of \state as a function of the average number of participant nucleons (rapidity) measured in Pb\--Pb collisions at $\sqrt{s_{\rm NN}}=2.76$ TeV.}
\label{fig-3}       
\end{figure}
  
The ALICE \state $R_{\rm AA}$~\cite{UpsiPbPbPN} measured at forward rapidity ($2.5<y_{\rm LAB}<4$) in Pb\--Pb collisions at $\sqrt{s_{\rm NN}}=2.76$ TeV is shown as a function of the number of participant nucleon in Fig.~\ref{fig-3} (left) and compared to that from the CMS Collaboration obtained at mid-rapidity ($|y_{\rm LAB}|<2.4$)~\cite{Chatrchyan:2012lxa}. In both experiments \state are measured down to $p_{\rm T}=0$. The suppression observed at forward rapidity is compatible with that at mid-rapidity for both central and semi-peripheral collisions. 

The ALICE and CMS $R_{\rm AA}$ as a function of rapidity are compared in Fig.~\ref{fig-3} (right). The \state suppression in the $2.5<y_{\rm LAB}<4$ rapidity range is compatible with that observed by the CMS Collaboration in the $|y_{\rm LAB}|<2.4$ range. 

Finally, the ALICE and CMS data suggest a rather small rapidity dependence for the \state suppression. It is to be noted that the the contribution of higher mass bottomonium to the observed suppression is not known and is expected to be large~\cite{Affolder:1999wm}.

\section{Conclusion}
The nuclear modification factor of \Jpsi has been measured down to $p_{\rm T}=0$ at forward ($2.5<y_{\rm LAB}<4$) and at mid-rapidity ($|y_{\rm LAB}|<0.9$) in Pb\--Pb collisions at $\sqrt{s_{NN}}=2.76$ TeV with the ALICE detector. A \Jpsi suppression is observed which is independent of the collision centrality for a number of participant nucleon larger than 100. Comparison of data with theory~\cite{Thermo,Zhao:2011cv,Liu:2009nb} suggests a large production of \Jpsi by (re-)generation in Pb\--Pb collisions at LHC energies.

The nuclear modification factor of \Jpsi has been measured down to $p_{\rm T}=0$ in p\--Pb collisions at $\sqrt{s_{\rm NN}}=5.02$ TeV in the system rapidity ranges $2.03<y_{\rm CM}<3.54$ and  $-4.46<y_{\rm CM}<-2.96$. The forward-backward ratio of $R_{\rm pA}$ has been also measured in the range $2.96<|y_{\rm CM}|<3.54$. The EPS09 calculations at Next-to-Leading-Order~\cite{Albacete:2013ei} reproduce reasonably the results, as well as models which include a contribution from coherent parton energy loss~\cite{Elossref}. On the contrary, the Color Glass Condensate Model~\cite{CGCref} is not favoured.

The \state \raa has been measured at forward rapidity ($2.5<y_{\rm LAB}<4$) and down to $p_{\rm T}=0$ in Pb\--Pb collisions at $\sqrt{s_{\rm NN}}=2.76$ TeV~\cite{UpsiPbPbPN}. A suppression of \state was observed. Comparison of ALICE data at forward rapidity and CMS data~\cite{Chatrchyan:2012lxa} at mid-rapidity ($|y_{\rm LAB}|<2.4$) suggests a rather small dependence of the suppression with rapidity. 

%
%

\end{document}